\documentclass[aps,prl,twocolumn,showpacs,superscriptaddress,10pt]{revtex4-1}

\usepackage{amsmath}
\usepackage{amsfonts}
\usepackage{amssymb}
\usepackage{graphicx}
\usepackage{color}
\usepackage{xcolor}

\newcommand{\beq}{\begin{equation}}
\newcommand{\eeq}{\end{equation}}
\newcommand{\bea}{\begin{eqnarray}}
\newcommand{\eea}{\end{eqnarray}}

\begin{document}
\title{A cavity-QED simulator of slow and fast scrambling}
%
%
\author{J. Marino}
\affiliation{JILA, NIST and University of Colorado, Department of Physics, University of Colorado, Boulder, CO 80309, USA}
\affiliation{Kavli Institute for Theoretical Physics, University of California, Santa Barbara, CA 93106-4030, USA}
\affiliation{Department of Physics, Harvard University, Cambridge MA 02138, United States}

\author{A. M. Rey}
\affiliation{JILA, NIST and University of Colorado, Department of Physics, University of Colorado, Boulder, CO 80309, USA}
\affiliation{Kavli Institute for Theoretical Physics, University of California, Santa Barbara, CA 93106-4030, USA}
\begin{abstract}
We study  information scrambling, as diagnosed by the out-of-time order correlations (OTOCs), in a system of  large spins collectively interacting  via  spatially inhomogeneous and incommensurate  exchange couplings. The model is realisable in a cavity QED system in the dispersive regime. Fast scrambling, signalled by an  exponential growth of the OTOCs, is observed when the  couplings do not factorise into the product of a pair of local interaction terms, and at the same time the state of the spins     points initially coplanar to the equator of the Bloch sphere. When one of these conditions is not realised, OTOCs grow  algebraically with an exponent sensitive to the orientation of the spins in the initial state. The impact of initial conditions on the scrambling dynamics is attributed to the presence of a global conserved quantity, which critically slows down the evolution for initial states close to the poles of the Bloch sphere.

%
%
\end{abstract}

\date{\today}
\maketitle

\emph{Introduction ---} Information scrambling, and its intimate relation to  quantum chaos and holographic duality~\cite{shenker2,maldacena,kitaev},   represents one of the frontiers of  research in condensed matter and many-body physics.
Out-of-time order correlations (OTOCs)
have been raised to the rank of prime quantifiers of information scrambling in this field:
for two commuting unitary operators $\hat{W}$ and $\hat{V}$, OTOCs are defined as
\begin{equation}\label{eq:otoc}
\mathcal{C}(t)=\langle [\hat{W}(t),\hat{V}(0)]^2 \rangle,
\end{equation}
with  $\hat{W}$  evolving with the Hamiltonian of the system, $H$.
~This quantity is currently considered  an indicator of the loss of memory of  initial conditions  in a quantum system.
Specifically, $\mathcal{C}(t)$ measures the overlap between two states: one state  is prepared through the subsequent application of  $\hat{V}$ at time $t=0$ and  $\hat{W}$ at a later time $t$, while the second   reversing this procedure in time.
%
\begin{figure}[t!]
\centering
\includegraphics[scale=0.3]{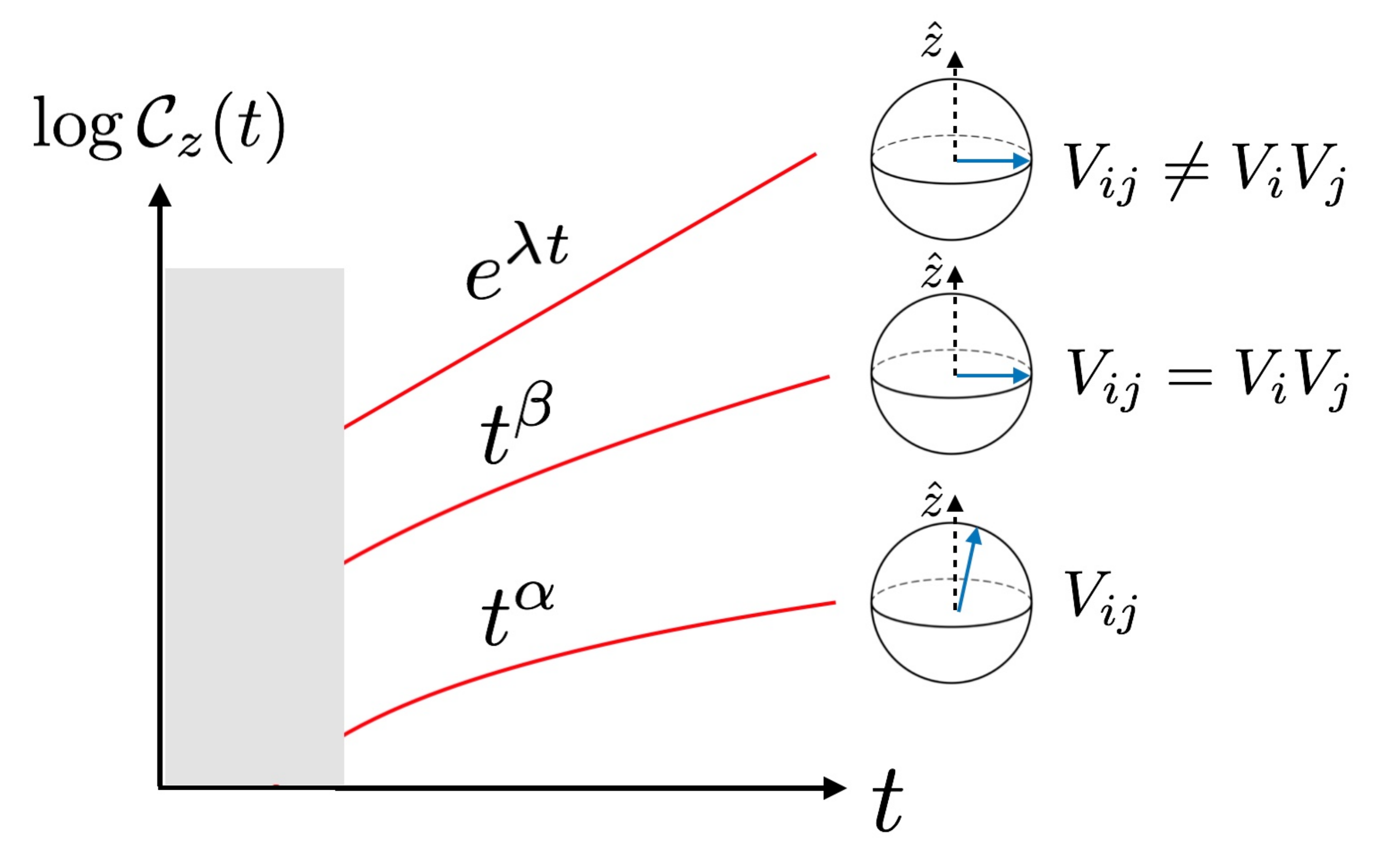}
\caption{ Sensitivity of the scrambling dynamics  to different initial conditions and to the nature of many-body coupling $V_{ij}$. For non-separable interactions $V_{ij}\neq V_iV_j$, and for initial spin states pointing close to the equator of the Bloch sphere, information is  scrambled fast in $\mathcal{C}_z(t)$, an OTOC computed from  $J^z_i$ spin operators (see Eq. \eqref{otocz}),  growing exponentially before saturation. In all other instances, the growth of $\mathcal{C}_z(t)$ is algebraic in time ($\alpha\simeq{5/2}$ and $\beta\simeq{9/2}$).}
\label{fig0}
\end{figure}

%

%
A system which deteriorates information exponentially in time is called a fast scrambler, and the associated scrambling rate, $\lambda$, has been regarded  as a quantum analog of the Lyapunov exponent,  which in classical chaotic systems dictates the rate at which two initially closed trajectories diverge in phase space~\cite{gutz}.  In quantum systems, $\lambda$,  as measured by the OTOC, is bounded, $\lambda\leq 2\pi T$ (with $T$ the temperature of the system), as firstly derived in the context of holographic theory and black holes~\cite{maldacena, susskind, shenker2}. 
However, in any system with finite  Hilbert space dimension, the growth in time of the OTOC  is a transient phenomenon before saturation to a constant value occurs, since, using triangular inequalities, the norm of $\mathcal{C}(t)$ in Eq.~\eqref{eq:otoc} can be always bounded by the norm of the operators therein involved.
Few condensed matter systems scramble fast~\cite{larkin, Das, ON, patel, szy, Alavirad, chav}, and, at the best of our current knowledge, none   saturates the bound on the quantum Lyapunov exponent, $\lambda$, with the exception  of the so called SYK model, a recent extension by Kitaev of an original model introduced by Sachdev and Ye~\cite{SY, parcol, sach, remark, altman2, Polchinski2016, bagrets2, Gross2017, Gross2017b, dries, ALTLAND201845,PhysRevB.95.205105, chen, garcia}, where information is scrambled   swiftly as in a black hole.
%
%
%
%
%
%
As  counterpart, slow scramblers are identified with those systems where out-of-time order correlations grow polynomially in time, with examples ranging from  interacting fermions in infinite dimensions~\cite{tsuji}, to Luttinger Liquids~\cite{moess},   many-body localised systems~\cite{he,Huse, Huang},  encompassing periodically kicked quantum Ising chains~\cite{kuk} and spin chains driven by noise~\cite{knap} (for a connection between scrambling and quantum critical points, see also~\cite{heyl,sahu}).
More recent developments include operator spreading and entanglement growth in quantum random circuits~\cite{keh, pollm, von, nahum, nah, chan, chan2, lamac,moud}, the search for velocity-dependent Lyapunov exponents in spatially-extended quantum systems~\cite{khemani, swingle, khemani2}, and a general  resurgence of interest in the concept of quantum chaos~\cite{faoro, tom, efim, liao}. Despite the  proliferation  of  theoretical studies  in the last few years, only few experimental proposals and realisations for the measurement of scrambling are currently available~\cite{meier}, encompassing ion traps~\cite{gart}, Ramsey interferometry~\cite{yao}, cavity QED systems~\cite{exp}, and nuclear magnetic resonance simulators~\cite{li, cap, niknam}.

 In this work, we consider a quantum many-body simulator of pairwise collectively interacting spins through position dependent couplings, a model realisable with atoms coupled to cavity photons, in the dispersive limit of the large cavity detuning and therefore capable to mediate interactions among the atoms. We  find that in our system the scrambling of  spin operators grows in a power law fashion, with the noticeable exception of exponential dynamics of the OTOCs  starting from spin states  pointing close to the $\hat{x}$-direction (or equivalently anyone coplanar with the equator of the Bloch sphere), and ruled by many-body interactions which do not factorize into pairs of local interaction couplings (see Fig.~\ref{fig0}  for a summary of  our results).
We use a  semi-classical treatment to study the scrambling dynamics, as done in several other models ranging from  kicked rotors~\cite{galit} to classical interacting spin chains~\cite{Das, moess2, silvia}. This is  motivated by  the observation that the chaotic behaviour  in the SYK model  is essentially of semi-classical nature as confirmed  in Refs. \cite{kurch, scafidi, dries}  which showed that quantum interference effects  renormalise the Lyapunov exponent to values consistent with the bound in Ref.~\cite{maldacena}.\\

\emph{The model ---}
We consider interacting particles of spin $L$  on the lattice, governed by the hamiltonian
\begin{equation}\label{model}
H=\frac{1}{{N{L}}}\sum^{{N}}_{i,j=1} V_{ij}~J^+_{i}J^-_{j},
\end{equation}
with $i=1,...,N$, and where $J^{\pm}_i$ are the raising/lowering operators of the $SU(2)$ algebra.
The pre-factors $1/N$ and $1/L$ are introduced  to render the Hamiltonian extensive in the thermodynamic limit, $H\propto N$, and scaling as $H\propto L$.
Rescaling the spin operators as  $\tilde{J}^{(\alpha)}_i\equiv{J}^{(\alpha)}_i/L$,
the commutation relations read ($\alpha$, $\beta$, $\gamma=x,y,z$; with $\tt{i}$ the imaginary unit)
\begin{equation}\label{eq:comm}
[\tilde{J}^{(\alpha)}_i,\tilde{J}^{(\beta)}_j]=\frac{\tt{i}}{L}\epsilon^{\alpha\beta\gamma}\delta_{ij}\tilde{J}^{(\gamma)}_j.
\end{equation}
Since $\tilde{J}^{(\alpha)}_i$ are bounded, Eq.~\eqref{eq:comm} implies that for large $L$, the Hamiltonian~\eqref{model}  describes the interaction among large, classical  spins.
  Note  that the magnitude of the spins is not controlled by the system size, rather by the independent parameter $L$.

The Hamiltonian~\eqref{model} can be realised for example considering a system of alkaline earth atoms exhibiting  a long-lived optical  transition, loaded in a cavity and tightly trapped in the ground vibrational level of a one dimensional  deep optical lattice, with lattice spacing  $\lambda_l/2$, and with $N_i$ atoms per site \cite{james1, james2,davis}.
In the dispersive regime  of large cavity detuning (see Supplemental Material (SM)), we can adiabatically eliminate the photons, and derive an effective Hamiltonian in the form~\eqref{model}.
We will consider: spatial dependent interactions which factorise, $V_{ij}=V_iV_j$ ('separable' interactions), with
\begin{equation}
\label{uno}
V_i=v\cos(2\pi i\sigma),
\end{equation}
and $\sigma=F_{M-1}/F_M$, where  $F_M\equiv N$ is the $M$-th term of the Fibonacci sequence; or
'non-separable' interactions ($V_{ij}\neq V_iV_j$), of the form
\begin{equation}
\label{due}
V_{i,j}=v^2\cos(2\pi (i-j) \sigma).
\end{equation}
This can be realised considering, in the former case, a  cavity which admits a single resonant mode propagating along the one-dimensional lattice with  wavevector $k = 2\pi/\lambda_c$, and, in the latter case,
 a ring cavity, which supports two degenerate running modes~\cite{Bux, sarang, Vaidya}. In both cases, we  assume that the ratio between $\lambda_c$ and   $\lambda_l$  incommensurate, and given by  $\sigma=\lambda_l/\lambda_c= 2 F_M/F_{M+1}$ (see SM for further details).\\

\emph{Out of time order correlations~(OTOCs) ---}
As customary in  spin systems with emerging classical behaviour, we introduce
$N$ pairs of canonical coordinates $(q_i,p_i)$ with $i=1,...,N$, representing, respectively, on the Bloch sphere,  the  azimuthal  angle, $0\leq q_i<2\pi$, and  the polar  angle, $0<\theta_i\leq\pi$, via the relation $p_i=\cos\theta_i$.
As anticipated, this constitutes an appropriate effective description of the degrees of freedom of the model, in the limit of a large number of atoms per lattice site ($N_i\simeq L\gg1$).
In turn, this allows to rewrite the normalised angular momentum components as
\begin{equation}\label{classici}
\tilde{J}^x_i=\cos q_i\sqrt{1-p^2_i},\quad \tilde{J}^y_i=\sin q_i\sqrt{1-p^2_i}, \quad \tilde{J}^z_i=p_i,
\end{equation}
and accordingly one can write the classicalised version of the OTOC  for the  $\hat{z}$-spin component
\begin{equation}
\label{otocz}
\begin{split}
\mathcal{C}_z&\equiv\langle[\tilde{J}^z_i(t),\tilde{J}^z_i(0)]^2\rangle=\\
&=\langle\{p_i(t),p_i(0)\}^2\rangle=\langle\left(\frac{\partial p_i(t)}{\partial q_i(0)}\right)^2\rangle.
\end{split}
\end{equation} 
%
%
%
%
\begin{figure}[t!]
\centering
\includegraphics[scale=0.25]{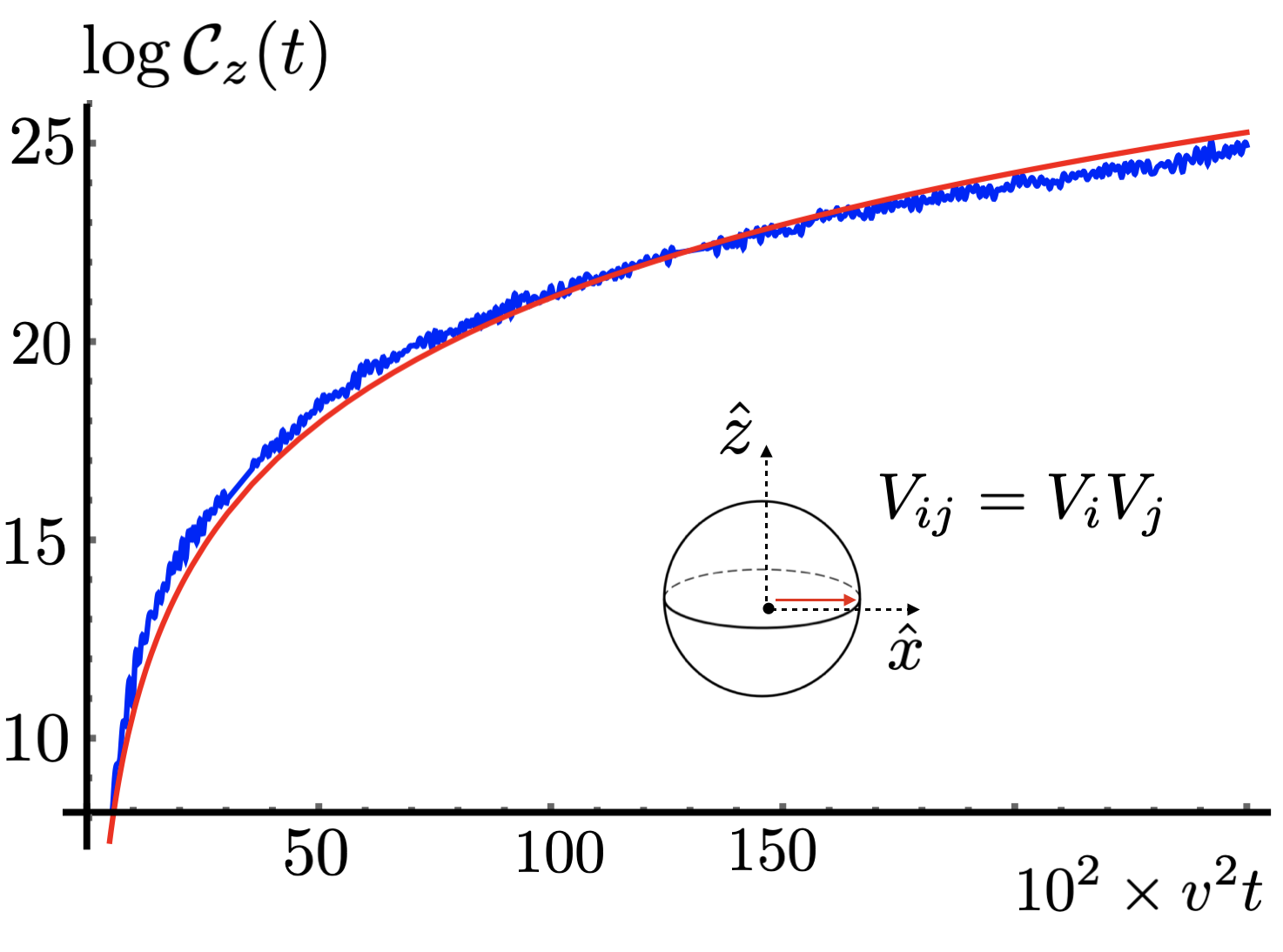}
\caption{Slow scrambling is signaled by the algebraic growth of $\mathcal{C}_z(t)$ (blue curve) as a function of time  for  separable interactions, $V_{ij}=V_iV_j$, and initial state pointing along the $\hat{x}$-direction. Here  $N=55$ sites ($N=F_{10}$) and $R=500$. The red curve is the semi-log fit, $\log \mathcal{C}_z(t)\simeq \beta \log (v^2t) + C$, with $\beta=4.55$ and $C=0.15$ (in the figure we plot the actual evolution over four decades; the logarithm is in natural basis). At longer times (not shown in the plot), $\mathcal{C}_z(t)$ saturates to a constant value. } 
\label{fig1}
\end{figure}
%
In Eq.~\eqref{otocz}, the quantum mechanical commutator has been converted into a Poisson parenthesis, given the emerging classical dynamics of \eqref{model}. Since the main focus of this work is on the slow and fast scrambling properties of the many body simulator in~\eqref{model}, we shall restrict our  calculations  for the rest of the paper on equal-site correlators for simplicity.

 The classical equations of motions for $(q_i,p_i)$  read
\begin{equation}\label{nonsep}
\begin{split}
\dot{p}_i=-&\frac{2}{N}\sqrt{1-p^2_i}\sum_{l}V_{il}\sin(q_i-q_l)\sqrt{1-p^2_l},\\
\dot{q}_i=&\frac{2}{N}\frac{p_i}{\sqrt{1-p^2_i}}\sum_lV_{il}\cos(q_i-q_l)\sqrt{1-p^2_l},
\end{split}
\end{equation}
as they can be straightforwardly derived from the Hamilton-Jacobi equations
\begin{equation}
\dot{p}_i=-\partial \mathcal{H}/\partial q_i, \quad \dot{q}_i=\partial \mathcal{H}/\partial p_i,
\end{equation}
with
\begin{equation}
\mathcal{H}=\frac{1}{N}\sum_{ij}V_{ij}\exp{(\tt{i}} {{(q_i-q_j)})}\sqrt{(1-p^2_i)(1-p^2_j)}.
\end{equation}
 the classical limit of $H$.
%
%
%
%

The instances of OTOC dynamics discussed in this work are realised starting from  a product state of identical coherent states  at  every lattice site, $|\psi\rangle_{(t=0)}=\prod^N_{i=1}\otimes |q_i,\theta_i\rangle $~(see also SM).
The state $ |q_i,\theta_i\rangle $ is   obtained rotating on the Bloch sphere the state $|L,-L\rangle_i$ on site $i$ by the pair of  angles $(q_i,\theta_i)$ (see the SM for the related expression), and, in the large $L$ limit, this corresponds to evolve the effectively classical dynamics of \eqref{model} from a set of initial random conditions drawn from a certain distribution, as it is done in semi-classical phase-space  methods~\cite{anatoli}.
%
%
%

\begin{figure}[t!]
\centering
\begin{tabular}{cc}
\includegraphics[scale=0.25]{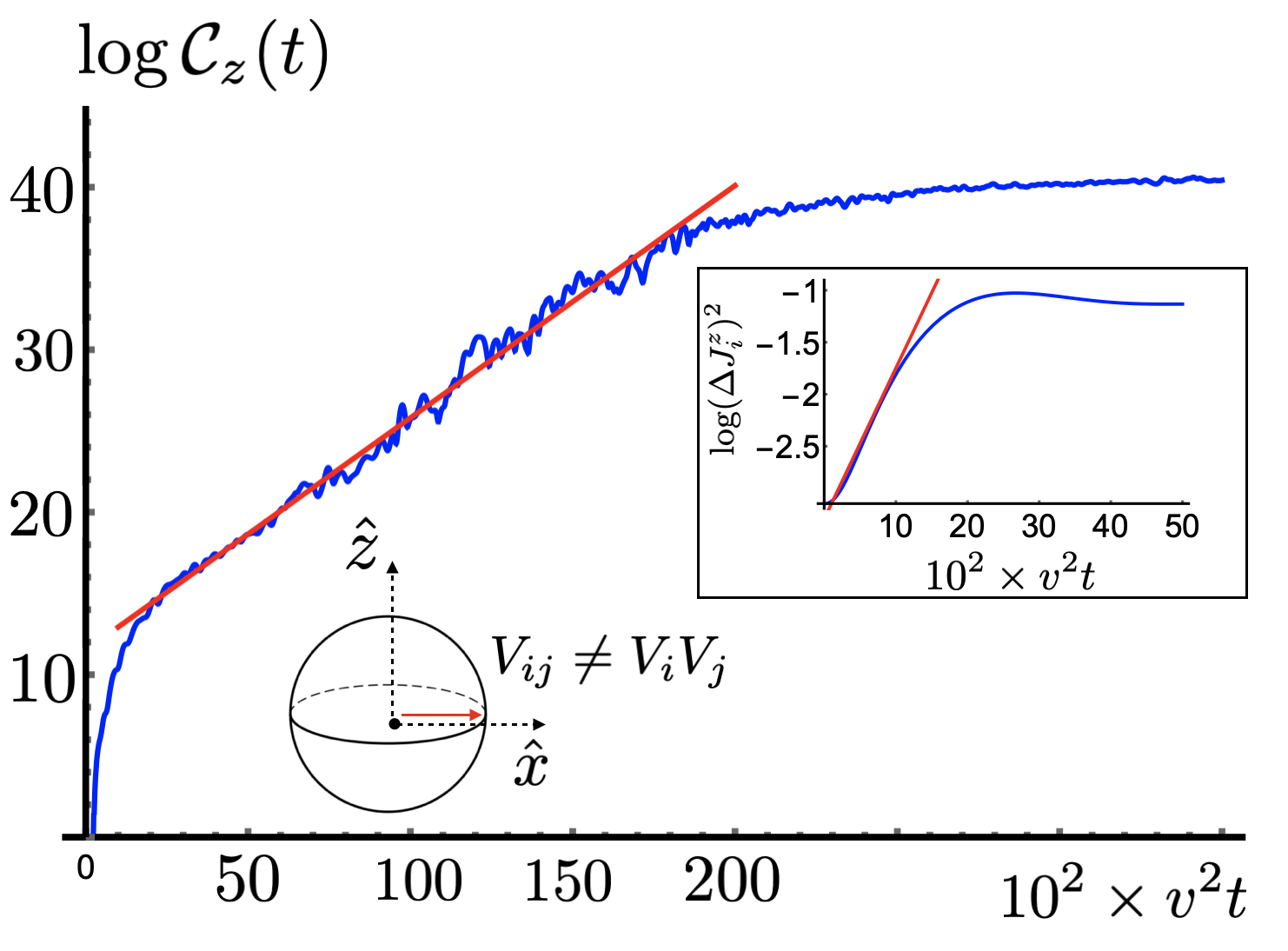}
\end{tabular}
\caption{Fast  scrambling is signaled by the exponential  growth of $\mathcal{C}_z(t)$ (blue curve) as a function of time  for  non-separable couplings, $V_{ij}\neq V_iV_j$, and initial state pointing along the $\hat{x}$-direction. Here  $N=55$ sites ($N=F_{10}$) and $R=500$. The red curve is the semi-log fit, $\log \mathcal{C}_z(t)\simeq \lambda v^2t + C$, with $\lambda=0.14$ and $C=11.5$ (in the figure we plot the actual evolution over three decades; the logarithm is in natural basis). At longer times  $\mathcal{C}_z(t)$ saturates to a constant value.  
\emph{Inset}: Semi-log plot of the variance $\log[(\Delta J^z_i)^2](t)\simeq  \lambda' v^2t+C'$, with $\lambda'=0.15$ and $C'=-3.3$.
}
\label{fig2}
\end{figure}
For instance, for initial states close to the north pole, one has (see Ref.~\cite{anatoli}) that $\langle J_i^z \rangle \simeq L$, $\langle J_i^{x,y} \rangle \simeq 0$ while ${\langle ({J_i^{x,y}})^2\rangle}={L/2}$.
 Following the semi-classical approach, we average over an ensemble of  $R$ different trajectories, each one resulting from a different realisation of the random initial condition.
Specifically, we  calculate  $\mathcal{C}_z(t)$ averaged over these $R$  classical trajectories, $ \mathcal{C}^a_z(t)$, with $a=1,...,R$, as
\begin{equation}
\mathcal{C}_z(t)=\frac{1}{R}\sum^R_{a=1} \mathcal{C}^a_z(t).
\end{equation} We have compared the classical dynamics sampled over this initial state distribution with a perturbative quantum mechanical calculation valid at short times, and found  quantitative agreement (see SM for details).\\

\emph{Slow scrambling ---}%
The OTOCs of the model show at short times  a  growth $\propto t^2$ as it can be shown with perturbation theory (see SM). However,  the non-linear character of interactions changes the time-dependent behaviour  of the OTOC  at later times.
In particular, we find that the choice of initial state and the separability of the interaction (or lack of it) influences the temporal dependence of the OTOC.
For initial states pointing close the poles of the Bloch sphere, the growth of $C_z(t)$ is power law  with $C_z(t)\propto t^\alpha$ for $\alpha\simeq2.5$ irrespectively of the separability of the interaction couplings, $V_{ij}$.
~Fig.~\ref{fig1} shows a second, qualitatively different instance of slow scrambling in the model:  $\mathcal{C}_z(t)\propto t^\beta$ with $\beta\simeq 4.5$,  for separable interactions and for an initial state pointing along the $\hat{x}$-direction on the Bloch sphere.\\ 

\emph{Fast scrambling ---} The non-separability of the exchange couplings, $V_{ij}\neq V_i V_j$, has  important consequences for initial states pointing along directions coplanar with the equator of the Bloch sphere.
In this case, the scaling  of the OTOCs at intermediate times is exponential, see Fig.~\ref{fig2}, indicating  that, in order to realise fast scrambling, the spatial structure of the interactions is crucial.
{Concerning the  late-time saturation  of the OTOCs, they both  reach the same asymptotic value in Figs. 2 and 3: different scrambling properties describe different ways for the OTOCs to relax, but the final steady state value  is  identical, provided one uses the same initial conditions  and   operators for the OTOCs.} %
%
%


\begin{figure}[t!]
\centering
\includegraphics[scale=0.25]{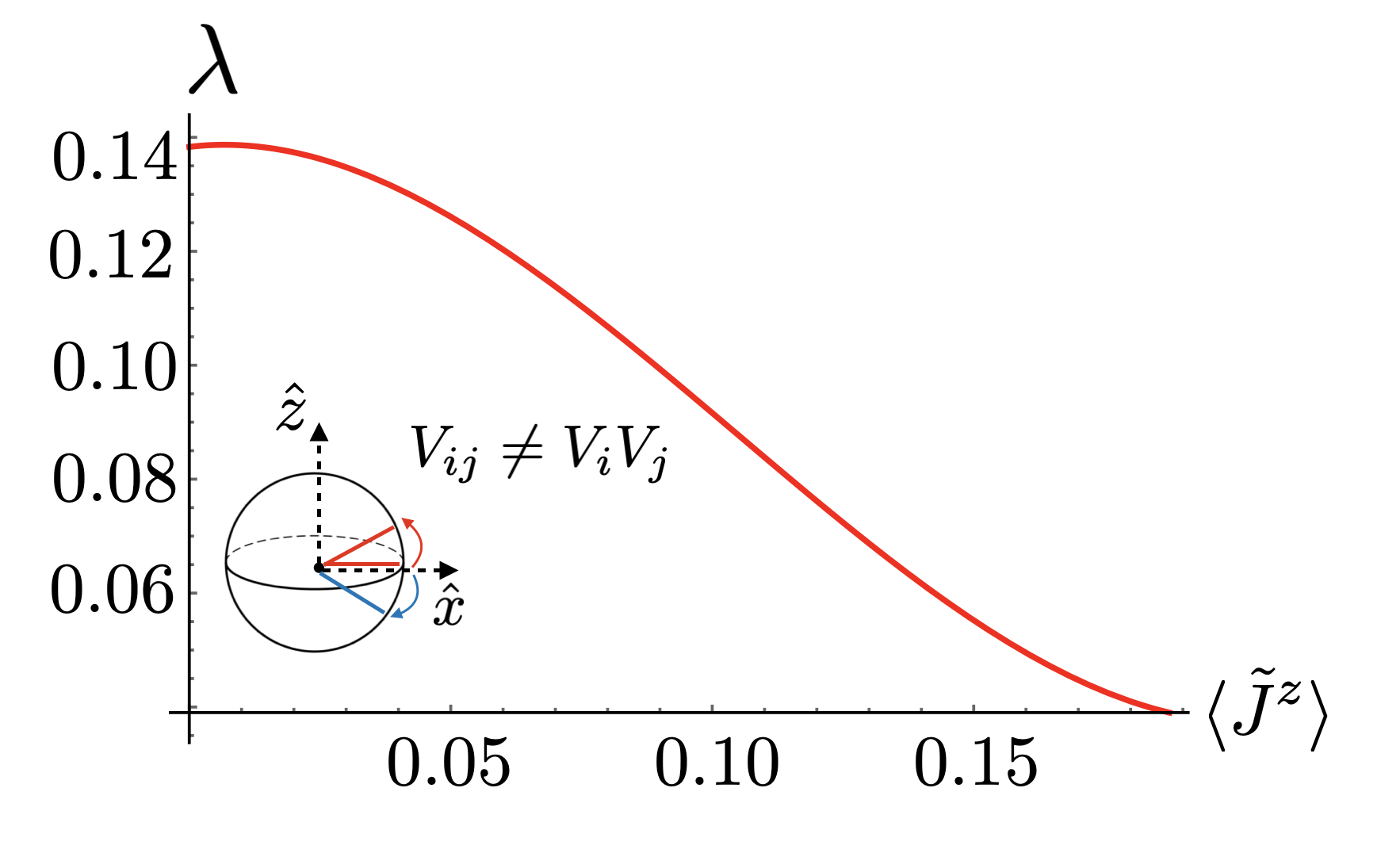}
\caption{Dependence of the scrambling  exponent, $\lambda$, as a function of the initial value of the rescaled magnetization $\langle {\tilde{{J}}^z}_i \rangle$ for non-separable  interactions. The red arrow on the Bloch sphere delimits the critical angle $\vartheta_c$, below which scrambling is slow (the blue line and arrow indicate that identical phenomenology holds in the southern hemisphere). $\lambda$ is extracted from $\mathcal{C}_z(t)$ evaluated on a system of  $N=55$ sites, with $R=300$.  }
\label{fig4}
\end{figure}
We now  consider  initial conditions  in the form of tensor products  of coherent states with every spin pointing in a direction tilted by a certain  polar angle $0<\vartheta<\pi/2$ with respect to the north pole of the Bloch sphere. Specifically, we want to study how the onset of fast scrambling in $\mathcal{C}_z(t)$ depends on $\vartheta$.
In Fig.~\ref{fig4} we report a specific instance  for a system of $N=55$ sites and  we find  that above  a certain critical angle, $\vartheta_c\simeq 0.44 \pi$, the out-of-time order correlator  $\mathcal{C}_z(t)$ still exhibits exponential growth before saturation with a Lyapunov exponent which vanishes continuously as $\vartheta\to\vartheta^+_c$.
Approximatively for $\vartheta\lesssim 0.3\pi$ slow scrambling $\propto t^{5/2}$ takes over again, while, in the between of these two critical angles, $\mathcal{C}_z(t)$ still grows  and saturates, although a neat fit is harder to find.
The presence of fast scrambling only for initial states pointing in a neighbourhood of the $\hat{x}$-direction, can be razionalised recalling that  $J^z\equiv \sum^N_{i=1}\hat{J}^z_i$, the total magnetisation along the $\hat{z}$-direction, is a conserved quantity in our model and  for states closer to the north pole  dynamics  becomes slower: in the limit case of maximal absolute value  of $\langle J^z \rangle$ (all spins pointing initially close to the north or south pole) no evolution occurs.
In Fig.~\ref{fig4}, we plot the dependence of the  exponent, $\lambda$, extracted from $\mathcal{C}_z(t)$, as a function of the  expectation value of $\tilde{J}^z_i$ (identical at each site, for the initial conditions we have chosen).
We have checked that for the next Fibonacci number (i.e. changing system's size), $N=F_{11}=89$, this scenario remains qualitatively unaltered.\\

\emph{Fidelity OTOCs ---} As recently pointed out in Refs.~\cite{robert}, the \emph{fidelity} OTOC, $\mathcal{C}_G(t)$, obtained setting in Eq.~\eqref{eq:otoc} the operator $\hat{V}=|\psi_0\rangle\langle \psi_0|$ equal to the projector on the state $|\psi_0\rangle$, and $\hat{W}=e^{i\delta\phi \hat{G}}$, with an Hermitian operator  $\hat{G}=J^z_i$ (in the case we analyse here), reduces for small perturbations, $\delta \phi\ll1$, to a measure of the variance, $\Delta G^2(t)$, of the observable $\hat{G}$. Specifically, up to second order in $\delta \phi$, we have~\cite{robert}
\begin{equation}\begin{split}\label{eq:otocf}&1-\mathcal{C}_G(t)=1-|\langle\psi_0|e^{i\delta\phi \hat{G}}|\psi_0\rangle|^2\simeq\\
&\simeq \delta\phi^2 (\langle G^2(t)\rangle -\langle G (t)\rangle^2)\equiv \delta\phi^2 \Delta G^2(t).
\end{split}\end{equation} The  \emph{fidelity} OTOC thus links fast scrambling  with the exponential grow
of $\Delta G^2(t)$,  and   represents  the  most  direct  diagnostic  of  chaotic
behaviour~\cite{gutz}. The importance of $\mathcal{C}_G(t)$ resides in the fact that variance is an easily accessible quantity  as demonstrated in trapped ion magnets~\cite{Bohnet1297} or cavity QED platforms~\cite{davis, leroux, chen, cox, Hosten, Hosten1552, lewis}.

%
We have calculated   the variance $ \Delta G^2(t)$ of initial states pointing along the $\hat{x}$ direction, in the scrambler given by~\eqref{model} with non-separable interactions, and we have found that it grows  as $ \Delta G^2(t)\propto e^{\lambda't}$, with $\lambda'\simeq0.15$, and saturates quickly,  in  agreement with the behaviour of $\mathcal{C}_z(t)$ (see also inset of Fig.~\ref{fig2}). The fact that the Lyapunov exponents extracted from the OTOC and the variance match, is consistent with the picture that  the onset of fast scrambling in our model is probably rooted in the presence of  an underlying classical chaotic regime for states pointing along the equator of the Bloch sphere.  \\

\emph{Perspectives ---}
In summary, our study presents an experimentally viable route for a system where the interplay of initial conditions and the structure of the many-body interactions  discriminate between the emergence of slow (algebraic in time) or fast (exponential) information scrambling in the OTOCs dynamics.
{ Notice  that upon reducing the number of atoms per lattice site, $L$, the system  enters a   regime dominated by quantum fluctuations. Although computations become more challenging, we don't expect that our main conclusions on  how separability of interactions gives rise to different scrambling properties  will  be modified; indeed,  in the case of  systems with reduced  $L$, the semi-classical dynamics of the OTOCs would most likely  represent a  saddle point solution which will be dressed by quantum corrections. One of the exciting aspects of our work is the possibility to test these theoretical  predictions, and resolve the 'quantum scrambling' properties of this model directly in the laboratory.}

The results  discussed here, open the way to the exciting perspective of studying the  scrambling properties of cavity QED simulators in the large cooperativity regime~\cite{monika, PhysRevLett33, colombe}, upon tuning the 'level' of randomness~\cite{pvt} in the spin-spin collective interactions, $V_{ij}$. For instance, it would be interesting to explore the impact of truly disordered interactions on  slow scrambling, or the robustness of fast scrambling when perfect periodicity is restored in  the spatial structure of the non-separable many-body interactions ({we don't expect that truly disordered couplings may compromise the fast scrambling behaviour, since this already occurs with the clean, yet irregular, distribution given by~\eqref{due}}).
~Furthermore, it would be worth to explore whether  slow and fast scrambling leave signatures on easily accessible observables through measurements of the output cavity field and collective spin distribution.

On the theory side, we believe it would be interesting  to study, in the future, the scrambling  of extended quantum systems with quasi-periodic (or quasi-random) short range many-body interactions, in order to explore how they would impact on the butterfly velocities  and front spreading properties of out-of-time order correlators.

\emph{Acknowledgements ---} We thank  E. Altman, J. Kurchan, A. Lerose,  B. Lev, R. Lewis-Swan, A. Polkovnikov, T. Prosen, A. Safavi-Naini for useful discussions. We thank R. M. Nandkishore and M. Norcia for comments on the manuscript. J.M. is supported by the European Union's Horizon 2020 research and innovation programme under the Marie Sklodowska-Curie grant agreement No 745608 (QUAKE4PRELIMAT). 
This work is supported by the Air Force Office of Scientific Research grants FA9550-18-1-0319 and its
Multidisciplinary University Research Initiative grant(MURI), by the Defense Advanced Research
Projects Agency (DARPA) and Army Research Office grant W911NF-16-1-0576, the National
Science Foundation grant  NSF PHY-1748958, PHY-1820885, JILA-NSF grant PFC-173400, and the National Institute
of Standards and Technology.

\bibliography{biblioSH}

\begin{widetext}
\newpage

\section{{Supplemental Material: \\ 
A cavity-QED simulator of slow and fast scrambling}}

\subsection*{Perturbation theory at short times}

In order to benchmark the semi-classical many-body dynamics  of the model (2), we  consider the short-time expansion for a generic operator $O(t)$, which reads
$O(t)\simeq O -$$\tt{i}$$t[O,H]-\frac{t^2}{2}[[O,H],H]+...$,
with $O$  the operator at time zero. We then expand accordingly the out-of-time order correlator $\mathcal{C}_z$.

For $J^z_i(t)$, this reads
\begin{equation}
\label{shorttime}
J^z_i(t)\simeq J^z_i-{\tt{i}}~t[J^z_i,H]+...
\end{equation}
which,  after some straightforward algebra, yields the following expression for the rescaled  angular momentum operator
\begin{equation}
\tilde{J}^z_i(t)\simeq \tilde{J}^z_i(0)-\frac{{\tt{i}}~t}{N}\left(\sum_b V_{ib}\tilde{J}^+_i\tilde{J}^-_b-\sum_a V_{a i}\tilde{J}^+_a\tilde{J}^-_i \right)+...
\end{equation}
and consequently an expansion for the OTOC in the form
\begin{equation}\label{otocper}
\mathcal{C}_z(t)=\langle\left[{J}^z_i(t),{J}^z_i(0)\right]^2\rangle\simeq\frac{t^2}{(LN)^2}\langle\psi| \sum_{b\neq i,b'\neq i}V_{ib}V_{ib'} (J^+_iJ^-_b+J^-_iJ^+_b)(J^+_iJ^-_{b'}+J^-_iJ^+_{b'}) |\psi\rangle.
\end{equation}
As discussed in the main text, in the large $L$ limit, the angular momentum operators have classical commutation relations (cf. Eq.~(3)), and the pairs of coordinates  $\{(q_i,p_i)\}_{i=1, ..., N}$ evolve according to  classical equations of motion, assuming one starts from a product state of coherent states. The expression of the latter  for a single spin reads
\begin{equation}\label{cohst}
|\theta,\varphi\rangle=\frac{1}{(1+|\tau|^2)^L}e^{\tau J_+}|L,-L\rangle=\sum^{L}_{m=-L}\frac{1}{(1+|\tau|^2)^L} {{2L}\choose{m+L}}^{1/2}\tau^{m+L}|L, m\rangle,
\end{equation}
where $\tau=e^{-i\varphi}\tan(\theta/2)$ and the index $m$ labels the eigenstates of $J_z$. The coherent state is  obtained with a  rotation of the state $|L,-L\rangle$,  respectively, by the polar and azimuthal angles $(\theta,\varphi)$. Accordingly, the state $|\psi\rangle$ in Eq.~\eqref{otocper} reads
\begin{equation}\label{prodotto}
|\psi\rangle=\prod^M_{i=1} |\theta,\varphi\rangle_i.
\end{equation}

We have benchmarked the short-time, perturbative, quantum mechanical calculation in Eq.~\eqref{otocper} and the semi-classical dynamics used in the main text for states pointing close to the north pole (analogous results hold for the south pole).
In this case, the magnitude of the angle involved in the rotation in Eq.~\eqref{cohst} can be deduced following the discussion contained in Ref.~[64]:  the    semi-classical spin state has variance along the $\hat{x}$ and $\hat{y}$ directions $\langle L_x^2\rangle=\langle L_y^2\rangle=L/2$, and therefore is closely aligned with the $\hat{z}$ direction. Specifically,  the angle measuring the departure from the $\hat{z}$-axis is given by $\tan\theta\sim1/\sqrt{L}$ and, for large $L$, this yields $\theta\sim1/\sqrt{L}$; in the notation of the canonically conjugated coordinates introduced in the main text, we have $p\simeq 1-1/(2L)$.
We have run the corresponding semi-classical dynamics, Eqs.~(8), and compared  with the quantum mechanical prediction at short times, Eq.~\eqref{cohst}, averaged over the state~\eqref{prodotto}. They both scales $\propto t^2$ and with  quantitative agreement  on the numerical pre-factor. \\

\subsection*{Details of the experimental implementation}
The model discussed in the  main text can be implemented using  electronic  long-lived optical  transitions in 
alkaline earth atoms  or hyperfine  ground state levels  in generic cold atom  laboratories via Raman transitions.  Here we focus on the former case but  generalization  to the latter can be straightforwardly carried out.  For the case of optical transitions, we  label the corresponding  electronic  levels as $g$ and  $e$ respectively. In addition we assume the atoms can have internal structure  consisting of  $\alpha=1,2,\dots, {\mathcal N}$  hyperfine levels. We consider the situation when  the atoms are tightly trapped in the ground vibrational level of a one dimensional  deep optical lattice along the $\hat{z}$ axis with lattice spacing  $\lambda_l/2$  inside of an optical  cavity. For simplicity, we assume the lattice  is at the  'magic wavelength' so that the trapping potential is the same for the different internal states. The Gaussian profile of the beams generates the transverse confinement which we approximate as an  harmonic potential where atoms  occupy the different transverse modes ${\bf n}=\{ n_x, n_y \}$.

We assume two different configurations for the cavity.
In the first case, the cavity admits a single resonant mode propagating along the one-dimensional lattice with associated wavevector $\vec{k}=k\hat{z}$ and  normalized mode profile of the form $\Xi(z)=\cos(k z)$. In the second case, we consider a ring cavity, which supports two degenerate running modes propagating along $\hat{z}$ with normalized mode profile $\Xi^\pm(z)\sim e^{\pm {\tt{i}} k z}$. In both cases  $k = 2\pi/\lambda_c$. We will also  assume that the ratio between $\lambda_c$ and   $\lambda_l$ is incommensurate, $\sigma=\lambda_l/\lambda_c= 2 F_{M-1}/F_{M}$ with $F_M$ the $M$-th term of the Fibonacci sequence, and
equal to number of lattice sites, $F_M\equiv N$.

The dynamics of the coupled  atom-light system is described by a master equation for the density matrix, $\hat \rho$,

\begin{equation}
 \frac{d\hat{\rho}}{dt} = -\frac{{\tt{i}}}{\hbar}\left[\hat{H}^{(1,R)},\hat{\rho}\right] + \mathcal{L}_c[\hat{\rho}]. \label{eqn:AL_master_eqn}
\end{equation}Here, the Hamiltonian describing the atom-light coupling is, for the single mode cavity

\begin{equation}\label{model1}
H^{(1)}=\Delta_c \hat{a}^\dagger \hat{a} +\sum_{\alpha=1}^{\mathcal{N}} \sum_{i=1}^{N} \sum_{{ \bf n}_{r_i}=1}^{N_i} g_{\alpha}\Xi(i)\Big (\hat{a}^\dagger {\hat \sigma}^-_{\alpha,i,{ \bf n}_{r_i}}+ \hat{a} {\hat \sigma}^+_{\alpha,i,{ \bf n}_{r_i}}\Big),
\end{equation} where $\hat{a} $ is the cavity mode annihilation operator, ${\hat \sigma}^+_{\alpha,i,{ \bf n}_{r_i}}\equiv |e,\alpha,i,{ \bf n}_{r_i}\rangle\langle   g,\alpha,i,{\bf n}_{r_i}|$ are Pauli raising operators for atoms in hyperfine state $\alpha$, lattice site $i$ and transverse mode ${\bf n}_{r_i}$.  $N_i$  is the number of atoms in lattice site $i$ and $\Xi(i)=\cos(2\pi\sigma i)$.

For the ring cavity, the Hamiltonian reads, instead,
\begin{equation}\label{model2}
H^{(R)}=\sum_{m=\pm1} \Delta_c \hat{a}_m^\dagger \hat{a}_m +\sum_{m=\pm 1}\sum_{\alpha=1}^{\mathcal{N}} \sum_{i=1}^{N} \sum_{{\bf n}_{r_i}=1}^{N_i} g_{\alpha}\Xi^\pm (i)\Big(\hat{a}_m^\dagger {\hat \sigma}^-_{\alpha,i,{\bf n}_{r_i}}+ \hat{a}_m {\hat \sigma}^+_{\alpha,i,{\bf n}_{r_i}}\Big).
\end{equation}Here $\hat{a}_\pm $ are  the corresponding cavity mode annihilation operators and $\Xi^\pm (i)=\exp[\pm {\tt{i}}    k i]$.

We have assumed   all the hyperfine levels are degenerate and  detuned from  the relevant cavity modes by $\Delta_c$ (there is no external magnetic field). In both cases the parameters  $g_{\alpha}$ give the dipolar couplings of the associated $|g,\alpha \rangle \leftrightarrow |e,\alpha\rangle $ transition (assuming linearly polarized light) and are proportional to the corresponding Clebsch-Gordan coefficient. We neglected any dependence on them on the  motional degrees of freedom, assumption  valid in the Lamb-Dicke regime.

The Lindblad term
\begin{equation}
 \mathcal{L}_c[\hat{\rho}] = \sum_m \frac{\kappa}{2}\left( 2\hat{a}_m \hat{\rho}\hat{a}_m^{\dagger} - \hat{a}_m^{\dagger}\hat{a}_m\hat{\rho} - \hat{\rho}\hat{a}_m^{\dagger}\hat{a}_m \right),
\end{equation}
describes photon loss from the cavity with  decay rate $\kappa$ (identical for all cavity  modes).

In the dispersive regime where $\Delta_c \gg \kappa, g_{\alpha}$, we can adiabatically eliminate the photons, obtaining
$\hat{a}_m (t)\approx  \frac{2}{2\Delta_c + {\tt{i}}\kappa} \sum_{\alpha=1}^{\mathcal{N}} \sum_{i=1}^{N} \sum_{{\bf n}_{r_i}=1}^{N_i} g_{\alpha}\Xi_m(i) {\hat \sigma}^-_{\alpha,i,{\bf n}_{r_i}}$.
By introducing the collective spin operators $\hat{J}^-_{\alpha,i}= \sum_{r_i=1}^{N_i} {\hat \sigma}^-_{\alpha,i,{\bf n}_{r_i}}$ Eq.~(\ref{eqn:AL_master_eqn}) simplifies the model  to a master equation for the reduced density matrix $\hat{\rho}_s$ of the spins,
\begin{equation}
 \frac{d\hat{\rho}_s}{dt} = -\frac{{\tt{i}}}{\hbar} \left[ \hat{H}_{\mathrm{eff}}^{(1,R)}, \hat{\rho}_s \right] + \hat{L}^{(1,R)}[\hat{\rho}_s]  \label{eqn:spin_master_eqn}.
\end{equation}
Here, the effective Hamiltonian for the one mode cavity is
\begin{equation}
 \hat{H}_{\mathrm{eff}}^{(1)} = \frac{4 \hbar    \Delta_c}{(4\Delta_c^2 + \kappa^2)}\sum_{j,l=1}^N \sum_{\alpha,\beta=1}^{\mathcal{N}} \cos(2\pi\sigma j)\cos(2\pi\sigma l) g_{\alpha} g_{\beta} \hat{J}^+_{j,\alpha} \hat{J}^-_{l,\beta}.
\end{equation}  This is accompanied by a
dissipative contribution which describes collective emission into the cavity mode,
\begin{eqnarray}
 \hat{L}^{(1)}[\hat{\rho}_s] = \frac{1}{2}\sum_{j,l=1}^N \sum_{\alpha,\beta=1}^{\mathcal{N}}\cos(2\pi\sigma j) \cos(2\pi\sigma l)\sqrt{\Gamma_{j,\alpha}\Gamma_{l,\beta}} \Big( 2\hat{J}^-_{j,\alpha}\hat{\rho}_s\hat{J}^+_{l,\beta}
  - \hat{J}^+_{j,\alpha}\hat{J}^-_{l,\beta}\hat{\rho}_s - \hat{\rho}_s\hat{J}^+_{j,\alpha}\hat{J}^-_{l,\beta} \Big),
\end{eqnarray}
where $\Gamma_{\alpha} = 4 \kappa g^2_{\alpha}  /(4\Delta_c^2 + \kappa^2)$.

For the ring cavity the corresponding Hamiltonian and Lindblad  terms are

\begin{equation}
\begin{split}
 \hat{H}_{\mathrm{eff}}^{(R)} &= \frac{4 \hbar    \Delta_c}{(4\Delta_c^2 + \kappa^2)}  \sum_{j,l=1}^N \sum_{\alpha,\beta=1}^{\mathcal{N}} \cos[2\pi\sigma (j- l)] g_{\alpha} g_{\beta} \hat{J}^+_{j,\alpha} \hat{J}^-_{l,\beta} ,\\
 \hat{L}^{(R)}[\hat{\rho}_s] &= \frac{1}{2}  \sum_{j,l=1}^N \sum_{\alpha,\beta=1}^{\mathcal{N}}  \cos[2\pi\sigma (j-l)]\sqrt{ \Gamma_{\alpha}\Gamma_{\beta}} \Big( 2\hat{J}^-_{j,\alpha}\hat{\rho}_s\hat{J}^+_{l,\beta}
  - \hat{J}^+_{j,\alpha}\hat{J}^-_{l,\beta}\hat{\rho}_s - \hat{\rho}_s\hat{J}^+_{j,\alpha}\hat{J}^-_{l,\beta} \Big).
  \end{split}
\end{equation}

In order to recover the hamiltonian in the main text we set $ \mathcal N=1$ (or $\mathcal N =2$, for hyperfine levels with the same $g_\alpha$), and we operate in the
 parameter regime  $\Delta_c\gg \kappa$, where  the Hamiltonian evolution  dominates over the dissipative terms (in the main text $\hbar=1$). 
 For  large number of atoms per site, $N_i\simeq L\gg1$, the spin operators $\hat{J}^+_i$ describe semi-classical operators of large angular momentum $L$, as discussed in the main text.\\

{ In current cavity  QED experiments  $\kappa$ is   typically of the order $100$ KHz. However, by improving the quality of the mirrors it is feasible to improve it by one or even two orders of magnitude. The cavity detuning from the
atomic transition frequency can be  easily varied from zero  to few  MHz.  Therefore the ratio   $\Delta_c/\kappa$ is highly tunable and can be made of $\sim 10$   easily in current experiments.}

{ In the paper, we have  normalized the Hamiltonian and thus the spin operators by the prefactor $L=N_i$, the number of atoms per lattice site. However in real experimental implementations this normalization  is not present making the dynamics $N_i$ times faster than the one shown in Figs. 2 and 3.
In the proposed implementation  the parameter $  N_i v^2 \sim  N_i C (\kappa/\Delta) \gamma$ is what controls the rate at which  the dynamics take place. Here $C= 4 g^2/(\kappa \gamma)$ is the single atom cooperativity,  and $\gamma$ is the single particle spontaneous emission rate, which can be the leading decoherence mechanism.}

{ As shown in the paper, fast scrambling can be seen when $N_i v^2t_s  \sim 50$.   Then for cooperativities $C\sim 10$ (such as the one reached already in Ref. [64]), $ N_i$ of the order of 1000,  $\Delta/\kappa\sim 10$, the scrambling time is  at the least an  order of magnitude smaller than the radiative decay.}\\

\subsection*{Expression for the OTOC of $\tilde{J}_i^x$}

The classical limit of the OTOC, $\mathcal{C}_x$, of $\tilde{J}_i^x$ reads

\begin{equation}
\label{otocx}
\mathcal{C}_x=\langle~~\left(\frac{\partial \tilde{J}^x_i(t)}{\partial q_i(0)}\right)^2\frac{\cos^2q_i(0)p^2_i(0)}{1-p^2_i(0)}+\left(\frac{\partial \tilde{J}^x_i(t)}{\partial p_i(0)}\right)^2\sin^2q_i(0)(1-p^2_i(0))-\sin (2q_i(0))p_i(0)\frac{\partial \tilde{J}^x_i(t)}{\partial q_i(0)}\frac{\partial \tilde{J}^x_i(t)}{\partial p_i(0)}~~\rangle
\end{equation}

where $q_i(0)$ and $p_i(0)$ are the initial values of the canonically conjugate coordinates of each angular momentum at site $i$,  $\tilde{J}^x_i$ is given by the first of Eq.~(6), and the average is taken over the same initial probability distribution used in the calculation of $\mathcal{C}_z$ and discussed in the main text.

\end{widetext}

\end{document}